\documentclass[a4paper]{jpconf}
\usepackage{graphicx}
\usepackage{amssymb}

\def\nc{\newcommand}

\nc{\shalf}{\ensuremath{\textstyle \frac{1}{2}}}
\nc{\deldag}{\mathbin{\partial\mkern-10.5mu\big/}}
\nc{\deldagss}{\mathbin{\partial\mkern-10.5mu/}}
\nc{\kdag}{\mathbin{k\mkern-10mu\big/}}
\nc{\udag}{\mathbin{u\mkern-10mu\big/}}
\nc{\kdagss}{\mathbin{k\mkern-10mu/}}
\nc{\Pdag}{\mathbin{P\mkern-10mu\big/}}
\nc{\pp}{{\scriptscriptstyle ||}}
\nc{\stwo}{{\scriptscriptstyle 2}}
\nc{\pham}{{\phantom{-}}}

\def\lsim{\mathrel{\raise.3ex\hbox{$<$\kern-.75em\lower1ex\hbox{$\sim$}}}}
\def\gsim{\mathrel{\raise.3ex\hbox{$>$\kern-.75em\lower1ex\hbox{$\sim$}}}}

\def\Slashnew#1{#1\kern-0.55em\raise.05ex\hbox{/}}
\def\slashnew#1{#1\kern-0.5em\raise.05ex\hbox{{$\scriptstyle /$}}}
\def\emph#1{{\em #1}}
\def\hepph#1{hep-ph/#1}

\nc{\beq} {\begin{equation}}
\nc{\eeq} {\end{equation}}
\nc{\beqa}{\begin{eqnarray}}
\nc{\eeqa}{\end{eqnarray}}

\def \wrt{{\em w.r.t.~}}
\def\prop{\Delta}
\def\sfrac#1#2{{\textstyle\frac#1#2}}

\begin{document}
\title{Coherent quasiparticle approximation (cQPA) and nonlocal coherence}

\author{Matti Herranen, \underline {Kimmo Kainulainen} and Pyry M. Rahkila}

\address{University of Jyv\"askyl\"a, Department of Physics,\\ 
        P.O.~Box 35 (YFL), FIN-40014 University of Jyv\"askyl\"a, Finland \\
        and \\
        Helsinki Institute of Physics, P.O.~Box 64, FIN-00014 University of  		
   	    Helsinki, Finland.}

\ead{kimmo.kainulainen@jyu.fi}

\begin{abstract} We show that the dynamical Wigner functions for noninteracting fermions and bosons can have complex singularity structures with a number of new solutions accompanying the usual mass-shell dispersion relations. These new shell solutions are shown to encode the information of the quantum coherence between particles and antiparticles, left and right moving chiral states and/or between different flavour states. Analogously to the usual derivation of the Boltzmann equation, we impose this extended phase space structure on the full interacting theory. This extension of the quasiparticle approximation gives rise to a self-consistent equation of motion for a density matrix that combines the quantum mechanical coherence evolution with a well defined collision integral giving rise to decoherence. Several applications of the method are given, for example to the coherent particle production, electroweak baryogenesis and study of decoherence and thermalization.
\end{abstract}

\section{Introduction}
\label{intro}

Many problems in modern particle physics and cosmology require setting up transport equations for relativistic quantum systems in out-of-equilibrium conditions, including baryogenesis~\cite{BG,ClassForce,ClassForceomat,SemiClassSK,PSW}, heavy ion collisions~\cite{baym} and out-of equilibrium particle production~\cite{brandenberger}. It is straightforward to write a formal solution for the problem in the Schwinger-Keldysh method, and detailed studies have been done on \emph{e.g.} thermalization of quantum systems~\cite{generic_therm}. It is much more difficult to find a simple enough an approximation scheme that is usable in practical applications. Here we will report a recent progress~\cite{HKR1,HKR2,HKR3,HKR4,HKR5} towards a scheme, in which one can account at least for certain particular types of  nonlocal coherence effects in the presence of hard collisions.  

The key observation is that in cases with particular symmetries (the homogeneous time dependent case and stationary, planar symmetric case) the free 2-point Wigner functions have new singular shell solutions (at $k_0=0$ for the homogeneous system and at $k_z=0$ in the static planar case), that carry the information about the nonlocal quantum coherence. When this shell structure is fed into the dynamical equations for the full interacting theory, they reduce to a closed set of quantum transport equations for the phase space distribution functions on the singular shells, which describe the on-shell particle numbers (mass-shells) and to a measure of the coherence (the new $k_{0,z}=0$ shells). 

\begin{figure}
\hspace{1pc}
\includegraphics[width=19pc]{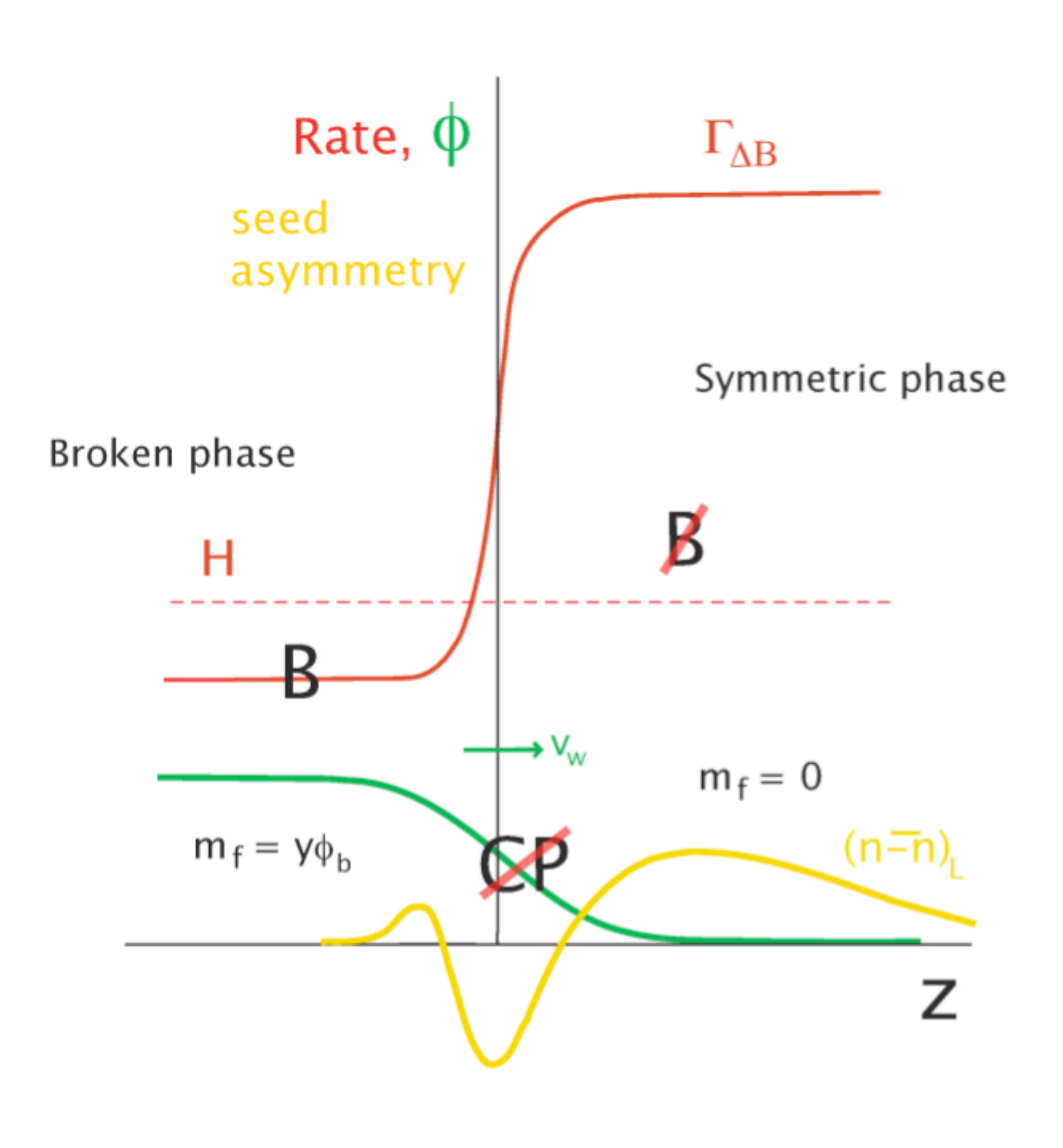}
\hspace{1pc}
\vspace{-1pc}
\begin{minipage}[b]{15pc}
 \caption{\label{fig:BG}
Shown are schematically how various quantities change over the profile of a planar phase transition front. The symmetric (massless) phase is to the right and the broken (massive) phase to the left, and the front is moving to the right with a velocity $v_w$.  The green line presents the variation of the Higgs field. The red solid line shows the baryon number violation rate $\Gamma_{\rm sph}$ and the red dashed line marks the constant expansion rate $H$. The solid yellow line shows the left chiral quark asymmetry.  \\\\\\}
\end{minipage}
\end{figure}

The problem that initially led us to develop the current formalism was the electroweak baryogenesis by the CP-violating quantum reflection mechanism. We therefore briefly discuss EWBG and the quantum reflection problem in this context. First, the baryon number is broken at nonperturbative level in the electroweak theory by the electroweak anomaly which couples the baryon and lepton currents to the Chern-Simons current in the gauge sector.  The anomalous B-number violation is vanishingly small in the zero temperature~\cite{thooft}, but it becomes very fast at high temperatures.  Above the {\em electroweak phase transition} temperature $T_c \approx 100$ GeV, where the electroweak symmetry is restored, the B-violation rate~\cite{RummuMoore} $\Gamma_{\rm sph} = (20-25) \alpha_w^5 T^4 \sim 10^{-6}T^4$, is much faster rate per degree of freedom than is the expansion rate of the universe during the transition: $H \sim 10^{-16}T$. Second, the P- and CP-parities, whose breaking is necessary to give direction to the B-violation, are broken by weak interactions themselves and by the complex phases in the quark mass matrix, respectively. CP-violating phases are also abundant (and needed) in many extensions of the standard model. A third condition for any successful BG-mechanism is to have out-of-equilibrium conditions. This is much harder to establish in the electroweak setting, because typical interaction rates between particles, $\Gamma_{\rm int} \sim \alpha_w^2 T \approx 10^{-3}T$, are enormous in comparison with the expansion rate during the transition. The only way to find out-of-equilibrium conditions with a standard expansion history is that the transition is of first order, proceeding through nucleation of broken symmetry (with $\phi\neq 0$, where $\phi$ is the vacuum expectation value of the Higgs field) bubbles inside the symmetric $\phi=0$ phase. In this case a new time scale, the bubble passing time, emerges: $\Delta t_b \approx l_w/v_w$, where $l_w$ is the width of the phase transition front and $v_w$ is the front velocity. Typically one finds $\Delta t_b \sim 10^{3}/T \sim \Gamma^{-1}_{\rm int}$. Because particles get their masses through the Higgs mechanism: $m \sim y \phi$, fermions change from being massless to massive excitations as they cross the phase transition front, and this sudden change is enough to push particle distributions momentarily out of equilibrium. Careful studies of the EWPT dynamics have shown that bubbles are created far apart and they grow to enormous dimensions before their coalescence completes the transition. Thus, in electroweak baryogenesis one is led to study the CP-violating, out-of-equilibrium interactions of fermions with expanding, stationary and planar phase transition walls. This setting is illustrated in Fig.~\ref{fig:BG}.

The precise quantitative problem is to compute the total left chiral quark asymmetry in the front of the transition wall, as this asymmetry acts as the seed that biases the B-violating rate to create the eventual baryon asymmetry.
This problem has been satisfactorily solved for the case of a very smooth phase transition front. This {\em semiclassical baryogenesis} formalism was first developed by use of WKB-techniques~\cite{ClassForceomat}, and later from a more fundamental CTP-formalism~\cite{SemiClassSK}. However, when the wall is not particularly thick, the quantum reflection mechanism may become relevant. This problem is much harder to solve in the presence of interactions because of the inherent nonlocality of the reflection mechanism. However, it can be addressed by use of our coherent quasiparticle approximation (cQPA) scheme, and solving this problem remains one of our immediate goals. However, there are also several other interesting applications for the cQPA formalism. These include a number of homogeneous problems, such as particle production by time-varying (oscillating) background fields, {\em e.g.}~during the preheating after inflation, or at the end of phase transitions. We have not yet completed implementing the cQPA scheme to the EWBG-program, but we have applied it to several homogeneous problems, some of which will be illustrated in the sections below.

This proceeding is organised as follows. In section \ref{sec:formalism} we introduce the basic CTP-formalism and introduce the Kadanoff-Baym equations in the mixed representation. In section \ref{sec:free} we derive the cQPA shell structure including the case with flavour mixing. In section \ref{sec:physical} we find expressions for physical quantities and solve some simple collisionless problems to illustrate the physical content of the coherence solutions. In section \ref{sec:interacting} we consider the interacting theory and derive the density matrix equation with collisions for the homogeneous problems. In section 
\ref{sec:scalar} we extend our results for the scalar fields, and finally section \ref{sec:conclu} contains our conclusions and outlook.

\section{CTP-formalism}
\label{sec:formalism}
\begin{figure}
\hspace{2pc}
\includegraphics[width=18pc]{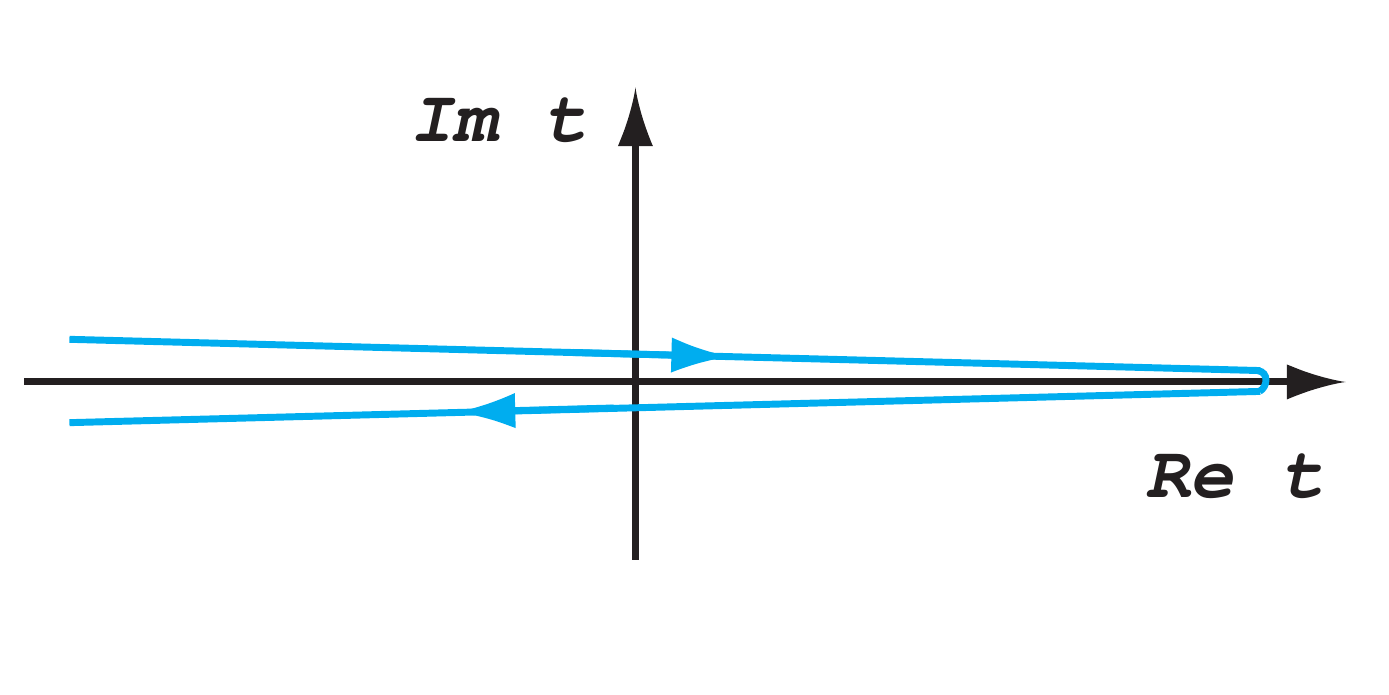}
\hspace{3pc}
\vspace{-1pc}
\begin{minipage}[b]{12pc}
 \caption{\label{fig:KeldyshPath}
          Schwinger-Keldysh path in complex time.\\\\\\}
\end{minipage}
\end{figure}

In this chapter we very briefly review the Schwinger-Keldysh formulation of a relativistic fermionic quantum field theory. The mathematical quantity of interest is the 2-point Wightmann function:
\beq
  iG^<_{\alpha\beta}(u,v) \equiv
     \langle  \bar \psi_{\beta}(v){\psi}_{\alpha}(u)\rangle \equiv
     {\rm Tr}\left\{\hat \rho \ \bar \psi_{\beta}(v){\psi}_{\alpha}(u)\right\}.
\label{G-less}
\eeq
By definition $G^<$ is an ensemble average over a quantum density operator $\hat \rho$, which fully describes the properties of the system. In the Schwinger-Dyson formalism one replaces the problem of finding $\hat \rho$ by a problem of finding all correlators defined by $\hat \rho$. Since $G^<$ is an  ``in-in''-correlator, a complex time contour has to be introduced, one convenient choice of which is shown in Fig.~\ref{fig:KeldyshPath}. On such a Closed Time Path (CTP)~\cite{Schwinger-Keldysh,CTP}, the basic object is a path ordered 2-point function (Dirac indices are suppressed): 
\beq
  iG_{\cal C}(u,v) =
            \left\langle T_{\cal C}
                 \left[\psi(u) \bar \psi (v)\right]
             \right\rangle ,
\label{Gcontour}
\eeq
where $T_{\cal C}$ defines time ordering along the Keldysh contour
${\cal C}$. The correlator $G_{\cal C}(x,y)$ can be shown to obey a contour Schwinger-Dyson equation
\beq
G_{\cal C} (u,v) = G^0_{\cal C} (u,v)
             + \int_{\cal C} {\rm d}^4z_1 \int_{\cal C}
                             {\rm d}^4z_2 \; G^0_{\cal C} (u,z_1)
               \Sigma_{\cal C} (z_1,z_2) G_{\cal C} (z_2,v)\,,
\label{SD1}
\eeq
where the self energy function $\Sigma_{\cal C}$ couples $G_{\cal C}$ to an infinite hierarchy of equations for higher order Green's functions. The precise form of $\Sigma_{\cal C}$ depends on the model Lagrangian and a truncation scheme for the hierarchy. (In the weak coupling limit a natural truncation is to substitute all higher than 2-point functions by their perturbative expressions.) %
The complex time valued function $G_{\cal C}(x,y)$ can be shown to obey a contour Schwinger-Dyson equation
The real time is regained at the expense of introducing a $2\times 2$ matrix structure; after some manipulations one can rewrite Eq.~(\ref{SD1}) in the following matrix form: 
\beq
G_0^{-1} \otimes G = \sigma_3 \; \delta +  \Sigma \otimes \sigma_3 G,
\label{SD3}
\eeq
where
\beq
G=\left(\begin{array}{cc}
            G_F & - G^< \\
            G^> & \phantom{-} G_{\bar F}
         \end{array}\right) \, , \qquad 
\Sigma=\left(\begin{array}{cc}
            \Sigma_F & - \Sigma^< \\
            \Sigma^> & \phantom{-} \Sigma_{\bar F} \,,
         \end{array}\right) 
\label{Gmatrix}
\eeq
and $\sigma_3$ is the usual Pauli matrix, and we defined a shorthand
notation $\otimes$ for the convolution integral:
\beq
    f \otimes g \equiv \int {\rm d}^4z f(u,z)g(z,v).
\label{otimes}
\eeq
Here $G_F$ and $G_{\bar F}$ are the chronological (Feynman) and
anti-chronological (anti-Feynman) Green's functions, respectively, and $G^<$
and $G^>$ are the Wightmann distribution functions, the former of which is just the correlator in Eq.~(\ref{G-less}). Above have also left out the labels $u$ and $v$ where obvious; for example $\delta \equiv \delta^4(u-v)$. Self energy functions can be calculated for example by using the 2PI effective action techniques (see eg.~\cite{kaksPI,CTP,PSW}). For example:
\beq
\Sigma^{<,>}(u,v) \equiv i \frac{\delta \Gamma_2[G]}{\delta
  G^{>,<}(v,u)} \,,
\label{2PIsigmas}
\eeq
where $\Gamma_2$ is the sum of all two particle irreducible vacuum graphs of the theory.  Introducing the usual retarded and advanced propagators $G^r(u,v) \equiv  \phantom{-} \theta(u^0-v^0) (G^< + G^>)$ and $G^a(u,v)\equiv -\theta(v^0-u^0) (G^< + G^>)$, and a similar decomposition for the self energy function $\Sigma$, equations (\ref{SD3}) can be recast into a particularly compact from:
\begin{eqnarray}
   (G_0^{-1}-\Sigma ^{r,a})\otimes G^{r,a} &=& \delta
\label{KB1a} \\
   (G_0^{-1}-\Sigma ^{r})  \otimes G^{<,>} &=& \Sigma ^{<,>}\otimes  
G^{a}.
\label{KB1b}
\end{eqnarray}
Equations (\ref{KB1a}) and (\ref{KB1b}) are called the pole equations and the \textit{Kadanoff-Baym equations}~\cite{KB}, respectively. In general, the former will fix the spectral properties of the theory, while the latter will give the dynamical evolution, including quantum transport effects we are interested in.
The KB-equation for $G^<$ can be further rewritten as
\beq
   (G_0^{-1}-\Sigma_H) \otimes G^< - \Sigma^< \otimes G_H
   = \frac{1}{2}\left( \Sigma^> \otimes G^< - \Sigma^< \otimes G^> \right) 
   \,,
\label{Dyneq}
\eeq
where the hermitian function $G_H\equiv \frac{1}{2}\left(G^r + G^a\right)$. The equation for $G^>$ is not needed, because $G^>$ and $G^<$ are related by the \textit{spectral function}, ${\cal A}\equiv \frac{i}{2}\left(G^r - G^a\right) = \frac{i}{2}(G^> + G^<)$, and ${\cal A}$ and $G_H$ are determined by the pole equations~(\ref{KB1a}). The spectral function is further subject to a condition $2 {\cal A}(t,\vec u; t, \vec v) \gamma^0 =  \delta^3(\vec u-\vec v )$, which is just the direct space version of the famous spectral sum-rule and follows directly from the canonical equal time anticommutation relation: $\{\psi(t,\vec u), \psi^\dagger (t,\vec v)\}  =  -i \delta^3(\vec u- \vec v)$.

\subsection{Mixed representation}

The next step, crucial to our program, is to move to the mixed representation Fourier transforming with respect to the relative coordinate $u-v$. However, it is necessary to first define more precisely the model we are interested in. With the electroweak baryogenesis and particle production problems in mind, we study a model described by the Lagrangian:
\beq
{\cal L} = i\bar \psi \deldag \psi
                    + \bar \psi_L m \psi_R
                    + \bar \psi_R m^* \psi_L + {\cal L}_{\rm
                      int} \,,
\label{freeLag1}
\eeq
where $m(x) = m_R(x) + im_I(x)$ is complex, possibly spacetime dependent mass and ${\cal L}_{\rm int}$ is the interaction part to be defined later. All nontrivial interactions responsible for coherence, will be produced by the mass term. From Eq.~(\ref{freeLag1}) it is easy to find that
\beq
G_0^{-1}(u,v) = \delta^4(u-v) (i \deldag_v - m^*(v) P_L - m(v)  
P_R)\,,
\label{free-prop}
\eeq
where $P_{L,R} = \frac{1}{2}(1 \mp \gamma^5)$ are the chiral projector matrices. Inserting the expression (\ref{free-prop}) into Eq.~(\ref{Dyneq}) and performing the Fourier (or Wigner) transformation we can rewrite the KB-equations in the mixed representation:
\beq
(\kdag + \sfrac{i}{2} \deldag_x - \hat m_0
       - i\hat m_5 \gamma^5) G^<
  -  e^{-i\Diamond}\{ \Sigma_H \}\{ G^< \}
  -  e^{-i\Diamond}\{ \Sigma^< \}\{ G_H \}
  = {\cal C}_{\rm coll},
\label{DynEqMix}
\eeq
where the $G$- and $\Sigma$ functions are functions of the average coordinate $x=\frac{1}{2}(u+v)$ and the $k$ which is the conjugate momentum to $r \equiv u-v$. For example the {\em Wigner} function $G^<(k,x)$ is defined as: 
\beq
G^<(k,x) \equiv \int d^{\,4} r \, e^{ik\cdot r}
                                         G^<(x + r/2,x-r/2) \,.
\label{wigner1}
\eeq
The collision term is given by
\beq
{\cal C}_{\rm coll} \equiv \sfrac{1}{2} e^{-i\Diamond}
                             \left( \{\Sigma^>\}\{G^<\} - \{\Sigma^<\}\{G^>\}
                             \right) \,.
\label{collintegral}
\eeq
The $\Diamond$-operator is the following generalization of the Poisson brackets:
\beq
\Diamond\{f\}\{g\} = \sfrac{1}{2}\left[
                   \partial_X f \cdot \partial_k g
                 - \partial_k f \cdot \partial_X g \right]
\label{diamond}
\eeq
and the mass operators $\hat m_0$ and $\hat m_5$ are defined as:
\beq
\hat m_{\rm 0,5} G^<(k,x) \equiv m_{R,I}(x) e^{-\frac{i}{2}
       \partial_x^m \cdot \partial_k^F} G^<(k,x)\,.
\label{massoperators}
\eeq
The KB-equation (\ref{DynEqMix}) together with the pole-equations (whose mixed representation forms we omit here) and the sum-rule form a complete set of equations for solving $G^<$, ${\cal A}$ and $G_H$ exactly, after the interactions, a scheme to compute $\Sigma$ and the mass profiles are specified. In practice these equations are way too hard to be solved in their full generality, and several approximations are needed to find a solvable set of equations. The novelty of this work is to show that a tractable approximation scheme that is general enough to treat quantum coherence simultaneously with interactions does exist.  We shall now proceed to build our scheme by first constructing the full phase space structure of the free 2-point functions.

\begin{figure}
\centering
\hskip -1truecm
\includegraphics[width=13cm]{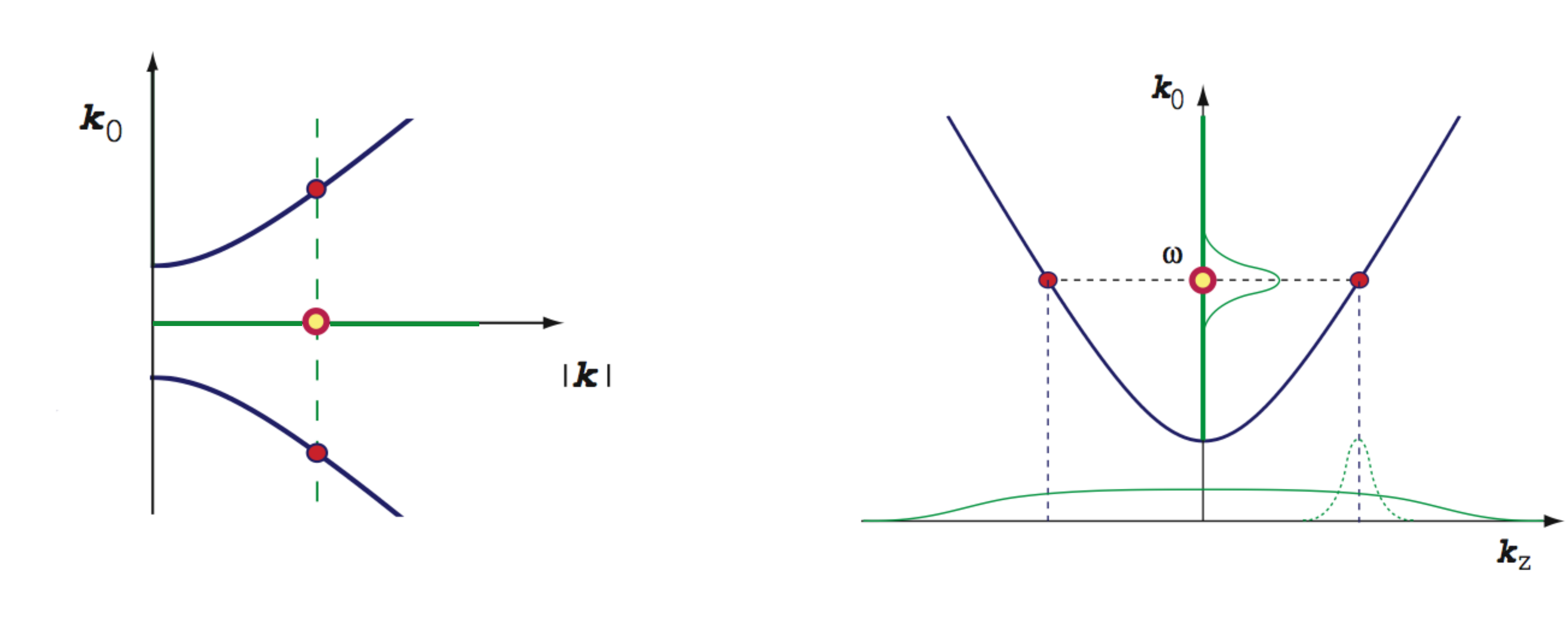}
\vskip -0.7cm 
    \caption{Dispersion relations in the case of a scalar mass function. 
    The dark filled (red) blobs show the mass-shell contributions and the light        (yellow) blobs show the corresponding coherence solution at the new $k_0=0$ shell in the homogeneous case (left) and at the $k_z=0$ shell in the planar symmetric case (right).}
    \label{fig:DR-omega}
\end{figure}

\section{Free theory and the shell structure}
\label{sec:free}

In the collisionless case the last three terms in equation (\ref{DynEqMix}) are absent. However, the mass operators still contain infinite number of momentum derivatives acting on $G^<$. These gradients are controlled by space-time derivatives of equally high order acting on the mass function. So, in the regions where the mass is slowly varying we can truncate these expansions to the lowest order. Note that the spatial variations in $G^<$ are not restricted by this approximation. In this limit we find
\beq
(\kdag + \sfrac{i}{2} \deldag_x - m_R - i m_I \gamma^5) G^< = 0 \,.
\label{simpleEq}
\eeq
Even this equation is very difficult to solve with general boundary conditions. We shall therefore restrict us to two special geometries relevant for the baryogenesis and particle productions problems: the spatially homogeneous and stationary planar symmetric cases. Consider first the homogeneous case. Observing that helicity is now a good quantum number, we can use the decomposition  %
\beq
iG_h^<\gamma^0  \equiv g_h^< \otimes \sfrac12(1 + h \hat k\cdot \vec \sigma)
\label{helicityansaz}
\eeq
in the chiral Weyl representation. When this decomposition is inserted to Eq.~(\ref{simpleEq}), it eventually breaks into hermitian and antihermitian 2x2-matrix equations for $g_h^<$:
\beq
{\rm (H)}:\quad 2 k_0 g_h^< = \{ H,g_h^< \}\,, \qquad
{\rm (AH)}:\quad i\partial_t g^<_h = [H, g^<_h] \,.
\label{Heq}
\eeq
where $H = - h |{\bf k}| \sigma_3 + m_R \sigma_1 - m_I \sigma_2$ is immediately identified as the the local Hamiltonian. The algebraic H-equation constrains the form of possible solutions. Using the Bloch representation: $g_h^< = \sfrac12(g^h_0 +  g^h_i \sigma^i)$, it can be written as $B_{ab}g^h_b = 0$, which has nontrivial solutions only if 
\beq
\det(B) = ( k_0^2 - {\bf k}^2 - |m|^2 )k_0^2 = 0 \,.
\eeq
Somewhat surprisingly, in addition to the usual mass-shell solutions there are new solutions living at shell $k_0=0$ (see Fig.~(\ref{fig:DR-omega})). The chiral structure of the mass-shell solution is
\beq
g^<_{h,{\rm m-s}}  = 
2 \pi s_0 \,f^h_{s_0} 
 \Big(k_0 - h|{\bf k}|\sigma_3 + m_R \sigma_1 - m_I\sigma_2 \Big) 
 \delta (k^2-|m^2|)  \,,
\label{geka}
\eeq
where $s_0 \equiv {\rm sgn}(k_0)$. This corresponds to the usual helicity eigenstate that can also be found by the use of Dirac equation. The chiral structure of the $k_0=0$ solutions is:
\beq
g^<_{h,{\rm 0-s}} =  \frac{\pi}{|{\bf k}|}
      \Big(h (m_R f^h_1 - m_I f^h_2)\sigma_3 
       + |{\bf k}|(f^h_1 \sigma_1 +  f^h_2 \sigma_2)\Big) \; \delta(k_0) \,,
\label{gtoka}
\eeq
Here the density functions $f^h_{\pm}$ and $f^h_{1,2}$ are some functions of time $t$ and the three momentum ${\bf k}$, whose evolution is determined by the dynamical equation. The mass-shell functions $f^h_{\pm}$ are related to the usual 1-particle densities of states and the new functions $f^h_{1,2}$ describe the quantum coherence between particles and antiparticles. The complete solution to the algebraic equation is compactly written as
\beq
g^<_h = 2\pi \left[ \sum_{\pm} \rho_{h\pm}\delta(k_0 \mp \omega_{\bf k}) +  \rho_{h0} \delta(k_0) \right] \,,
\label{gkolmas}
\eeq
where $\omega_{\bf k}^2 \equiv {\bf k}^2+|m^2|$ and the matrices $\rho_{h\pm}$ and $\rho_0$ can be read off from Eqs.~(\ref{geka}) and (\ref{gtoka}). The full shell structure of the homogeneous model is shown in the left panel of the Fig.~\ref{fig:DR-omega}.

A similar construction can be done for the stationary, planar symmetric case~\cite{HKR1}. In that case the spin $s$ orthogonal to the symmetry plane is conserved and eventually, in a particular frame~\cite{HKR1}, one finds equations very similar to Eqs.~(\ref{Heq}):
\beq
-2 s k_z g^<_s = P g_s^< + g_s^< P^\dagger \,, \qquad
is \partial_z g^<_s = P g_s^< - g_s^< P^\dagger \, ,
\label{Peq}
\eeq
where $P \equiv k_0 \sigma_3  - i(m_R \sigma_2 + m_I\sigma_1)$ is the local  momentum operator. The algebraic equation again gives rise to a singular shell structure; only this time the mass shell is accompanied by a singular momentum shell at $k_z=0$, where $k_z$ is the momentum perpendicular to the planar symmetry in the static rest frame. We do not write the explicit forms for these solutions here, but they can be found in ref.~\cite{HKR1}. The complete shell structure for the correlator in the planar symmetric case in is shown in the right panel of the Fig.~\ref{fig:DR-omega}. 

The crucial issue leading to the new solutions in either of the cases discussed above is that \emph{we did not assume translational invariance}.  Indeed, if one imposes translational invariance (in time in homogeneous and in space in the planar symmetric case), the dynamical equations in (\ref{Heq}) and (\ref{Peq}) become algebraic and it is then easy to show that they exclude the $k_{0,z}=0$ solutions.  Another important thing to observe is that the new $k_{0,z}=0$ solutions do not appear in the spectral function despite the fact that in the collisionless limit the equation for the spectral function coincides with Eq.~(\ref{simpleEq}) for $G^<$.  The difference comes from the additional constraints from the spectral sum rule which are enough to kill the coherence solutions from ${\cal A}$. Thus, as one might expect, coherence solutions exist only as correlations between particular 1-particle phase space states in the dynamical sector of the theory.

\subsection{Flavour mixing}

\begin{figure}[h]
\includegraphics[width=22pc]{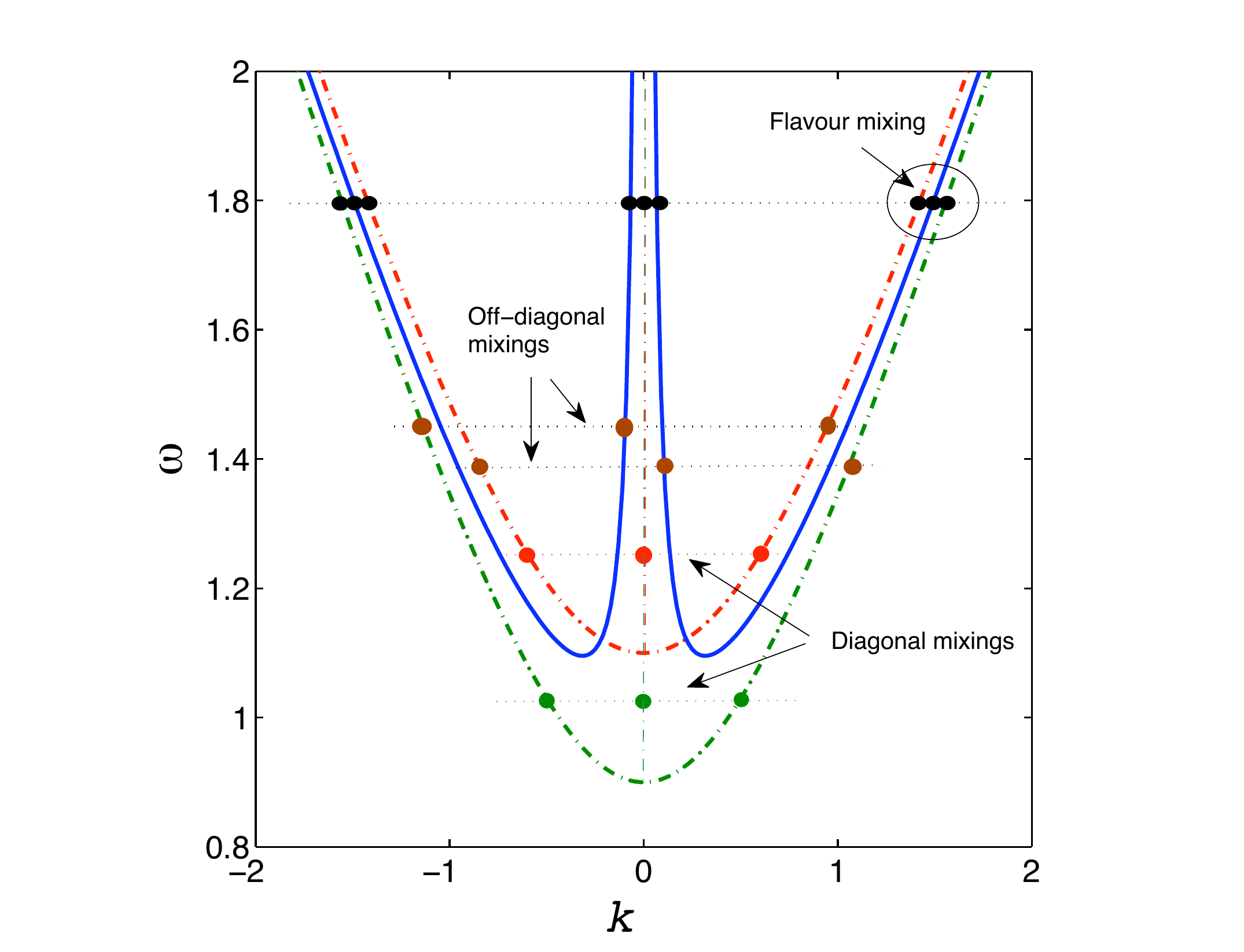}
\hspace{0pc}%
\vspace{-1pc}
\begin{minipage}[b]{15pc}
 \caption{\label{fig:flavourmix}
          Shown is the complete set of dispersion relations corresponding to   
          the case of $2\times 2$ flavour mixing. The triplets of filled circles connected by dotted lines indicate states which at a given energy form the chiral mixing sub-complexes. The black filled circles indicate a possible flavour mixing complex when the chirality mixing is prohibited by extrenous constraints.\\\\}
\end{minipage}
\end{figure}

The discussion of this section can be straightforwardly extended to the case of several mixing fields. We shall consider here only the planar symmetric case. If we replace the simple complex mass $m$ by a general complex mass matrix $M$, the analysis leading to equations~(\ref{Peq}) remains intact, apart from the fact that the local momentum operator becomes a matrix: $P \equiv k_0 \sigma_3  - i(M_H \sigma_2 + M_{AH}\sigma_1)$, where $M \equiv M_H + i M_{AH}$. The matrix structure makes extracting the spectral solutions more complicated. However, for a hermitian $M$ the dispersion relation can still be expressed in a very simple form:
\beq
(k_z)_{ij} = \pm \frac{1}{2} \left( \sqrt{k_0^2 - m_i^2} \pm \sqrt{k_0^2 - m_j^2} \right) \,,
\eeq
where $m_{i,j}$ refer to the mass eigenstates of $M$. This dispersion relation shows new interesting features.  The diagonal entries obviously correspond to the simple DR's found above in the scalar mass case. The off-diagonal entries however correspond to the singular shells for the {\em flavour mixing} coherence. Interestingly, the latter DR's change continuously from one closely degenerate with mass shells and describing mixing between particles going mostly in the same direction to a shell close to $k_{z}=0$ and describing the combined flavour mixing and reflection coherence between particles going on opposite directions. The complete phase space structure is shown in Fig.~\ref{fig:flavourmix} for the case of a hermitian $2\times 2$ mixing matrix. Clearly a very complicated set of mixing scenarios is possible with this dispersion relation complex, where one can have the two diagonal 1-particle densities coupling up to eight coherence functions. On the other hand, if the extrenous constraints are enough to restrict the mixing to, say, positive momenta only, but not enough to resolve the flavour mixing triplet  (illustrated within the black circle in Fig.~\ref{fig:flavourmix}), one finds the usual flavour oscillation pattern widely studied in neutrino physics.

The complex singular shell structure illustrated in this section is the foundation of, and gives the name to our approximation scheme: the coherent quasiparticle approximation or cQPA.

\section{Physical density matrix}
\label{sec:physical}

The dynamical equations in (\ref{Heq}) and (\ref{Peq}) look compelling. The time dependent equation (\ref{Heq}) in particular has just the usual form of the  evolution equation for a density matrix. However, they are actually meaningless because of the singular structure imposed by the algebraic constraints. That is,   the $g^<_{h,s}$-matrices are mere phase space densities that cannot be identified with physical density matrices. As was already suggested in the previous section, we will have to convolve them with some $k_0$- and/or $k_z$-dependent weight functions that may quantify the prior extrenous information on the system~\cite{HKR1}. Different types of such information functions were illustrated in the right panel of Fig.~\ref{fig:DR-omega} by dashed lines. A {\em physically observable} density matrix can be formally written as
\beq
  \rho_{\cal W} (k_0,k_z,s;z)
     \equiv \sum_{s'} \int \frac{{\rm d}k'_0}{2\pi} \frac{{\rm d}k'_z}{2\pi}\;
      {\cal W} (k_0,k_z,s \mid\hskip-0.5truemm\mid k'_0,k'_z,s'\,;\,z)
      \; g^<_{s'}(k_z',k_0';z) \,,
\label{rhoW}
\eeq
where ${\cal W}$ encodes the prior information. Let us make clear that this description is an effective one; in fact all restrictions should be imposed on the system directly by the collision integral. For example, our knowledge that a certain neutrino beam is moving in a given direction should be described by the time-dependent correlations imposed on the system by interactions with the apparatus that generates the beam. However, it is obvious that the effective description can often be more useful in practice.

In many applications some parameters can be fixed by conservation laws while no prior information at all exists on some other parameters. For example, in our reflection problem the energy, the momentum parallel to the symmetry plane and the spin $s$ are conserved, while we cannot assume any information on the perpendicular momentum of the state. The weight function appropriate for describing such a system then is ${\cal W} = 2\pi \delta(\omega - k_0' ) \, \delta_{s,s'}$.  Using this weight in Eq.~(\ref{rhoW}) one finds that the density matrix is just the zeroth moment of the fundamental distribution with respect to $k_z$: $\rho_{\cal W} \equiv \rho_s (\omega,s) = \langle g^<_s\rangle$. Note that in general there is no guarantee that the equation of motion for a weighted density matrix has a simple form, but in this particular case, because the momentum operator $P$ depends only on $k_0$, the collisionless equation for $\rho_s$ indeed is just
\beq
is\partial_z \rho_s = P \rho_s - \rho_s P^\dagger \,. 
\label{Peq2}
\eeq
Unlike Eq.~(\ref{Peq}), the new evolution equation (\ref{Peq2}) is well defined because the matrix $\rho_s$ is a nonsingular object that depends only on the energy $\omega$ and spin $s$.  A similar construction can be made in the context of a homogeneous problem, where one can fix the momentum $\bf k$ and helicity, and integrate over all energies. The resulting collisionless equation, now for a density matrix $\rho_W \equiv \rho_h(|{\bf k}|, h)$, is found to be the familiar one:
\beq
i\partial_t \rho_h = [H, \rho_h ] \,,
\label{Heq2}
\eeq
where $\rho_h = \rho_{h+} + \rho_{h-} + \rho_0$ in terms of the matrices defined in equation (\ref{gkolmas}).

\subsection{Moment connections}

From practical point of view the most important thing about the singular shell structure we have found is that it allows one to draw a one-to-one correspondence between the density matrix elements and the shell $f_i$-functions. Indeed, hermitian $\rho$ matrices contain four degrees of freedom, which is exactly the number of independent shell functions. Clearly, if the coherence shell solutions were ignored two degrees of freedom would be missing, making it impossible to describe coherence phenomena. Using the Bloch-representation $\rho_h \equiv \frac{1}{2}(\langle g^h_0 \rangle + \langle g^h_i \rangle \sigma^i)$, we have the following relations between $\langle g^h_\alpha \rangle$ (here the zeroth moments \wrt energy)  and $f^h_{\pm,1,2}$:
\begin{eqnarray}
\langle g^h_0 \rangle &=& \phantom{-}f^h_+ + f^h_-
\nonumber \\ [0.5mm]
\langle g^h_1 \rangle &=& \phantom{-}\sfrac{{m_R}}{\omega}
    (f^h_+ - f^h_-) + f^h_1
\nonumber \\[0.5mm]
\langle g^h_2 \rangle &=& -\sfrac{{m_I}}{\omega} (f^h_+ - f^h_-) + f^h_2
\nonumber \\
\langle g^h_3 \rangle &=& -h\sfrac{{|{\bf k}|}}{\omega} ( f^h_+ - f^h_-) 
+ h \Big(\sfrac{{m_R}}{{|{\bf k}|}} f^h_1 - \sfrac{{m_I}}{{|{\bf k}|}}f^h_2\Big) \,,
\label{rhocompbloch}
\end{eqnarray}
where $\omega \equiv \omega_{\bf k}$. As we shall see later, these moment-$f_i$ function relations will allow one to rewrite the collision integral entirely in terms of the moment functions, leading to a solvable, closed system of equations for the density matrix. 

\subsection{Physical quantities}

It is gratifying to see that the natural definition of the number density in terms of our mass-shell density functions coincides with some other nonequilibrium definitions found in the literature. Indeed, according to the usual Feynman-Stuckelberg interpretation the phase-space particle number density is identified as $n \equiv f_+$, while for (fermionic) antiparticles $\bar n \equiv 1 - f_-$. With these identifications, using the inverse relations of Eq.~(\ref{rhocompbloch}) we find that:
\beqa
n_{{\bf k}h} &=& \frac{1}{2 \omega}\left(-h|{\bf k}| \langle
  g^h_3 \rangle + m_R \langle g^h_1 \rangle - m_I \langle g^h_2
  \rangle \right) + \frac{1}{2} \langle g^h_0 \rangle 
\nonumber \\ 
{\bar n}_{{\bf k}h} &=& \frac{1}{2 \omega}\left(-h|{\bf k}| \langle
  g^h_3 \rangle + m_R \langle g^h_1 \rangle - m_I \langle g^h_2
  \rangle \right) - \frac{1}{2} \langle g^h_0 \rangle + 1 \,.
\label{partnumber}
\eeqa
Setting ${\rm Tr}(\rho^h) = \langle g^h_0 \rangle \equiv 1$ (physically this constraint corresponds to setting the chemical potential to zero) these expressions reduce to the ones obtained in 	ref.~\cite{Prokopec_partnumber}\footnote{Different signs of terms involving $h$ and $m_I$ are due to a different sign convention in our definition of the Hermitian Wightmann function.}, where they were derived using the solutions to a Dirac equation and a Bogolyubov transformation to diagonalize the fermionic Hamiltonian. 

One can also compute the energy and the pressure densities from the diagonal elements of the energy moment tensor $\theta^{\mu\nu}=\frac{i}{4}(\bar\Psi\gamma^\mu\partial^\nu\Psi  
- \partial^\nu\bar\Psi\gamma^\mu\Psi) + \mu \leftrightarrow \nu $. For energy density one finds
\beq
\langle{\cal H}(x)\rangle = \langle\theta^{00}(x)\rangle 
= \sum_h \int \frac{{\rm d}^3 k}{(2\pi)^3}\,\omega_{\vec
  k}\left(n_{{\bf k}h}+{\bar n}_{{\bf k}h} -1\right)\,,
\label{energy_dens}
\eeq
which, according to expectations is simply a sum of free particle and antiparticle contributions, plus the vacuum energy. For the pressure we find
\beq
\langle P(x)\rangle = \langle\theta^{ii}(x)\rangle 
= \sum_h \int \frac{{\rm d}^3 k}{(2\pi)^3}
\,\frac{1}{3}\left(\frac{{{\bf k}}^2}{\omega}\left(n_{{\bf k}h}+{\bar
      n}_{{\bf k}h}  - 1\right) - m_R f^h_1 + m_I f^h_2 \right) \,. 
\label{pressure}
\eeq
Now, in addition to normal free particle and antiparticle terms there is an explicit contribution from the coherence shell functions $f_{1,2}$. Thus the pressure is different from the statistical one at the quantum level. However,  the coherence functions $f_{1,2}$ are oscillatory with microscopic time-scales $\Delta t_{\rm osc} \sim 1/\omega$ and the  classical thermodynamical pressure arises when Eq.~(\ref{pressure}) is averaged out over time-scales exceeding the quantum scale $\Delta t_{\rm osc}$.

\subsection{Solutions: collisionless examples}

We applied Eq.~(\ref{Peq2}) to the Klein problem of reflection off a step potential wall in ref.~\cite{HKR1}. Correct results were found for the transmission and reflection coefficients and tunneling factors when the coherence shell factors are included in Eqs.~(\ref{rhocompbloch}), while neglecting them gave only pure classical complete transmission and complete reflection solutions. Similar results were later found also for the bosonic Klein problem~\cite{HKR2}.
\begin{figure}[h]
\vspace{-2pc}
\hspace{-2pc}%
\includegraphics[width=22pc]{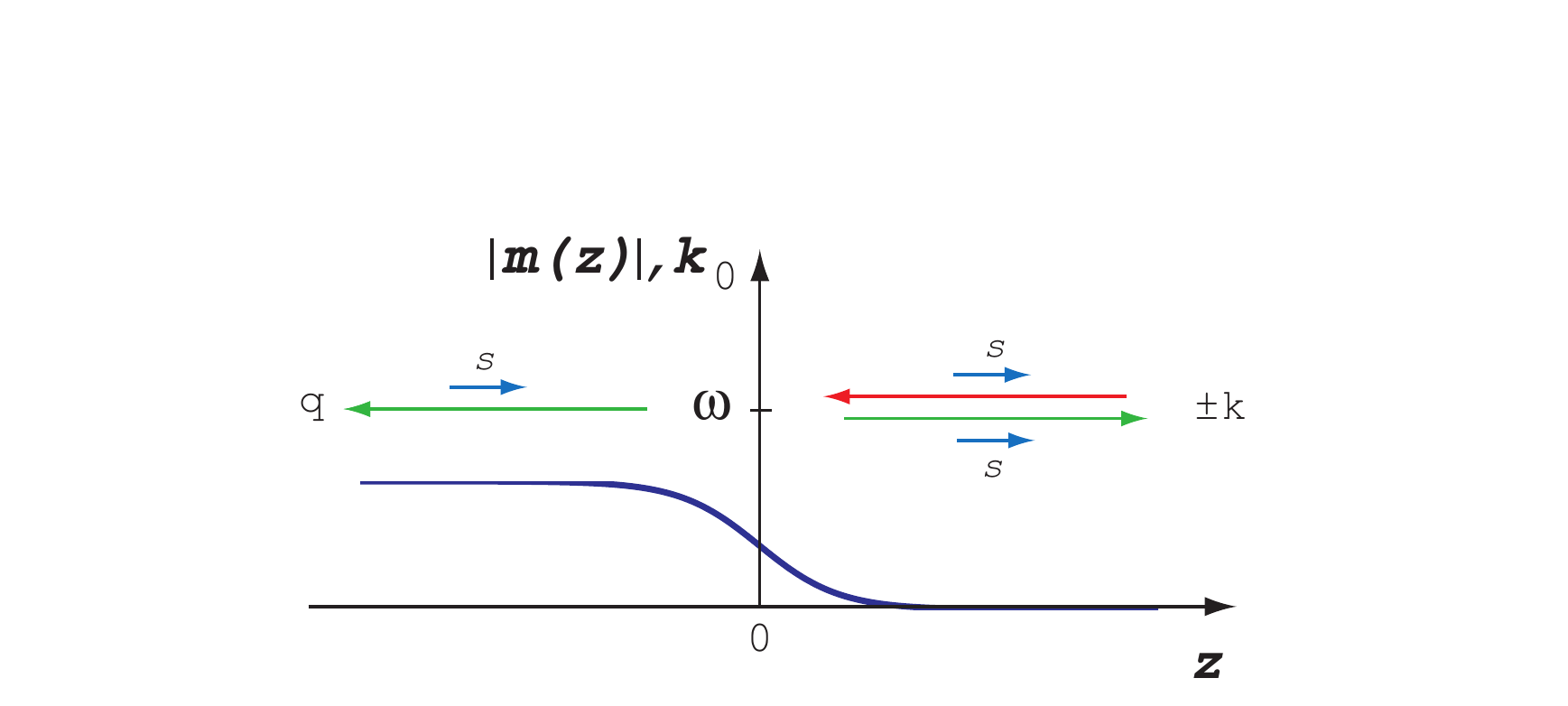}
\hspace{1pc}%
\vspace{0pc}
\begin{minipage}[b]{15pc}
 \caption{\label{fig:reflection}
          Reflection from a potential wall due to spatially varying complex mass function.\\\\}
\end{minipage}
\end{figure}
We also considered a reflection off a smooth mass wall, shown in Fig.~\ref{fig:flavourmix}. This is precisely the problem of interest for the quantum reflection mechanism in the electroweak baryogenesis. Again, the integrated moment equations are exact for this case, and the moment connections are needed only asymptotically to define the reflection coefficients.
\begin{figure}[h]
\vspace{1pc}
\hspace{0pc}
\includegraphics[width=16pc]{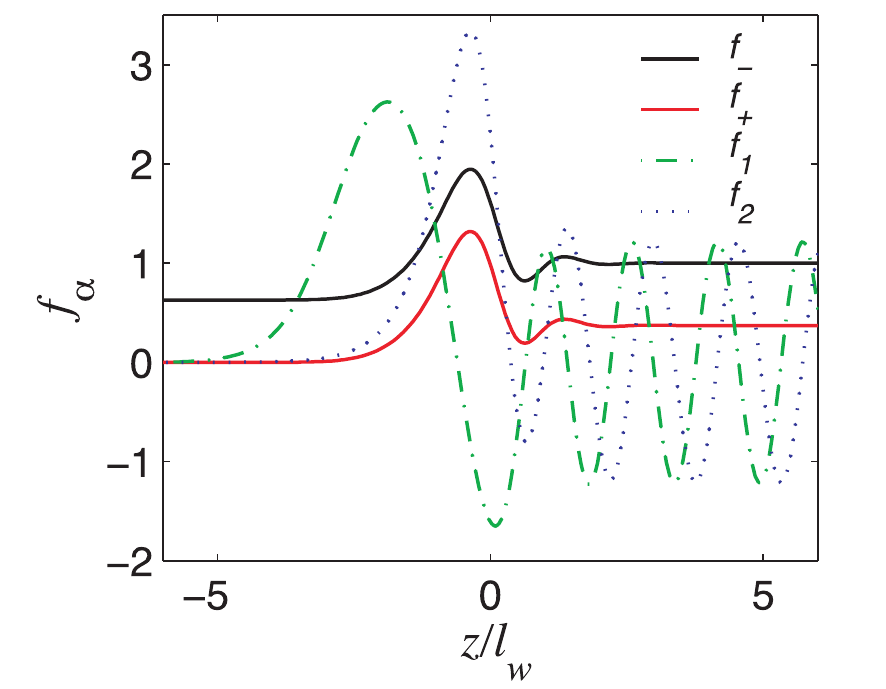}
\hspace{5pc}%
\vspace{-1pc}
\begin{minipage}[b]{15pc}
 \caption{\label{fig:deltaj}Shown are the mass-shell functions $f_\pm$       
             corresponding to the
             left and right moving fluxes and functions $f_{1,2}$, which 
             encode the quantum coherence. We have taken $s=1$ and 	 	
             $q/|m_{-\infty}|=0.088$.\\\\\\\\}
\end{minipage}
\end{figure}
The solutions for the mass-shell functions $f_-$ and $f_+$ (here equivalent to the left and right moving currents, respectively) and the coherence functions $f_{1,2}$ are shown in figure~\ref{fig:deltaj}.  The coherence functions are oscillating in the massless phase and vanish in the massive phase. This is as expected because the boundary conditions (see Fig.~\ref{fig:flavourmix}) exclude right moving states at $z\rightarrow -\infty$. Our results for the reflection and transmission currents agree with those found earlier with use of the Dirac equation~\cite{CKV}. The crucial improvement over these results is to include the effects of interactions. We have not yet completed this work for the planar symmetric case. However, parallel results for the homogeneous problem do exist and they will be presented next.

\section{Interacting fields}
\label{sec:interacting}

With interactions included the full Schwinger-Keldysh equations become much more complicated of course. A tractable approximation scheme can still be obtained by combining the familiar quasiparticle approximation with our zeroth gradient order treatment of the constraint equations. In this scheme we can in particular drop~\cite{HKR2} the $G_H$-term in Eq.~(\ref{DynEqMix}). Eventually we find:
\beq
(\kdag + \sfrac{i}{2} \deldag_x - \hat m_0
       - i\hat m_5 \gamma^5 - \Sigma_H ) G^<
  = {\cal C}_{\rm coll} \,.
\label{DynEqMix2}
\eeq
Here we are focusing only on the effects of the nontrivial mass function, so we
drop the thermal corrections to the dispersion relation setting $\Sigma_H=0$. After inserting the helicity decomposition Eq.~(\ref{helicityansaz}), breaking the equation to the hermitian and antihermitian parts and, consistently with our cQPA scheme, dropping the collision integral in the constraint equation, which is also expanded to the lowest order in gradients\footnote{Note that these are just the standard assumptions that one makes when deriving the usual Boltzmann equation from the Kadanoff-Baym equations. See {\em e.g.}~\cite{Henning}.}, we find:
\beqa
{\rm (H):} && \quad 2k_0 g^<_h = \{H, g^<_h\} 
\nonumber \\
{\rm (AH):} &&\quad \partial_t g^<_h = -i[H, g^<_h] - {\cal C}_h^- \,,
\label{rho_collHOMOG2}
\eeqa
where ${\cal C}_h^-$ is the antisymmetric part of the helicity projected collision integral. In cQPA and for an uncorrelated (for example a thermal) self-energy it can be written as~\cite{HKR5}
\beq
i{\cal C}_h^- = \cos(\sfrac{1}{2}\partial^\Gamma_{k_0}\partial^g_t) \{\Gamma_h,g^<_h\}
+ i\sin(\sfrac{1}{2}\partial^\Gamma_{k_0}\partial^g_t) [\Gamma_h,g^<_h ]
- \{\Gamma_h, g^<_{\rm eq}\} \,,
\eeq
where $\Gamma_h \equiv \sfrac{i}{2}(\Sigma_h^> + \Sigma_h^<)$ and we defined the hermitian helicity projected self energy function $P_h\gamma^0\Sigma^{<,>}P_h \equiv \Sigma^{<,>}_h \otimes \sfrac12(1 + h \hat k\cdot \vec \sigma)$. Finally $g_{\rm eq}^<$ is the equilibrium limit of the distribution $g^<_h$. We were careful to keep a subset of gradients to infinite order, because the projection onto $k_0$ shell of these expansions is not controlled by any small parameters ($\Gamma$ or $m'$) and need to be resummed. Of course equation (\ref{rho_collHOMOG2}) is equally badly defined as was our free theory evolution equation for $g^<$. It is also clear that when the integration is done, the integrated collision term will, thanks to the moment connections Eq.~(\ref{rhocompbloch}), reduce to and expression entirely expressible in terms of the density matrix components. When integration (over $k_0$ with a flat weight) and resummation is done, one eventually finds~\cite{HKR5}:
\beq
\partial_t \rho_h = -i[H, \rho_h]
- \{\Gamma_{hs}, \delta \rho_h \}
- \{\Gamma_{ha}, \delta \rho_{h+} - \delta \rho_{h-}\}
- \frac{1}{2\omega_{\bf k}} [\Gamma_{ha}, [H,\rho_h]]\,.
\label{rho_coll_intHOMOG}
\eeq
Here brackets $\{,\}$ denote an anticommutator,  $\rho_h = \rho_{h+} + \rho_{h-} + \rho_0$ in terms of the matrices in Eq.~(\ref{gkolmas}) and $\delta \rho_i \equiv \rho_i - \rho_{i,{\rm eq}}$, and the collision terms $\Gamma_{h(s,a)}\equiv \frac{1}{2} (\Gamma_{h+} \pm \Gamma_{h-})$ denote symmetric and antisymmetric combinations of the mass-shell interaction strengths $\Gamma_{h\pm}\equiv \Gamma_h(k_0 = \pm\omega_{\bf k})$.  Equation (\ref{rho_coll_intHOMOG}) is the sought after generalization of the standard collisionless equation of motion (\ref{Heq2}) for the density matrix to the case with decohering collisions.  

\begin{figure}
\centering
\includegraphics[width=9cm]{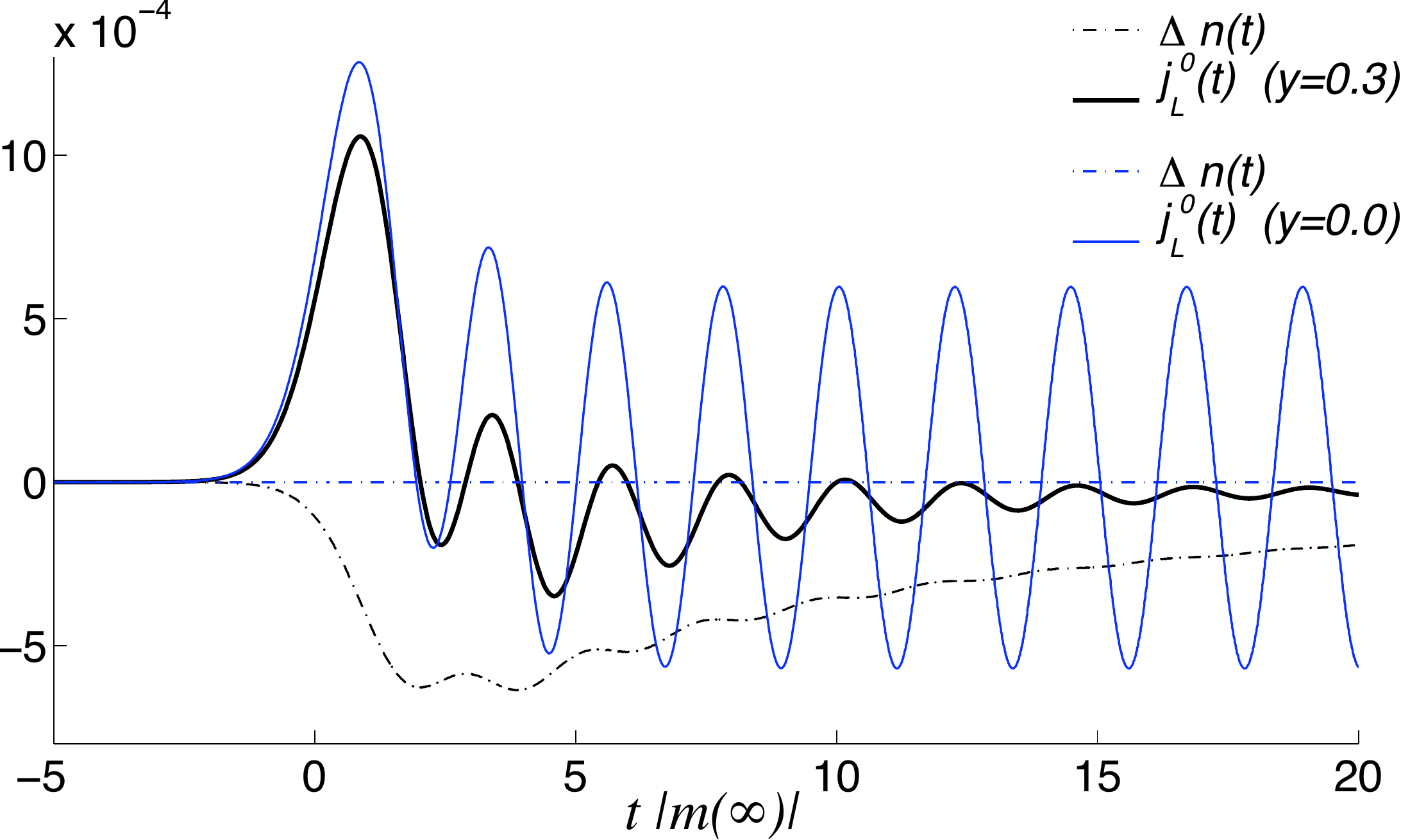}
\vskip -0cm 
    \caption{Shown are the total asymmetry 
    $\Delta(n) = \sum_{h} (n_{{\bf k}h}-\bar{n}_{{\bf k}h})$ 
    (see  Eq.~(\ref{partnumber})) and the L-chiral current density per unit 
    ${\bf k}$ volume: 
    $j^0_{L{\bf k}}(t) =  \sum_h ({\rm Tr}[\frac12(1+\sigma_3)\rho_h]- \sfrac12)$ (vacuum part subtracted). Blue curves correspond to the noninteracting and black curves to the interacting cases.}
    \label{fig:fermions}	
\end{figure}

We have applied the equation of motion (\ref{rho_coll_intHOMOG}) to study a system with a temporally evolving $CP$-violating mass function:
\beq
|m(t)| = \frac{1}{2}\big(1+{\rm tanh}(t/\tau_w)\big)\,,
\qquad
\arg[m(t)] = \frac{1}{2}\Delta\theta \big(1-{\rm tanh}(t/\tau_w)\big) \,,
\label{phiprof}
\eeq
where $\tau_w$ is the characteristic time of the transition. Collisions were modelled by an interaction with a thermally equilibrated fermion and scalar fields $q_R$ and $\phi$: 
\beq 
{\cal L}_{\rm int} = - y\; \bar \psi_L \phi \, q_R + h.c. 
\label{interaction}
\eeq
The lowest order contribution to the self energy from this interaction is:
\beq \Sigma^{<,>}(k,x) = i |y|^2 \int
\frac{d^4k'}{(2\pi)^4} P_R G_q^{<,>}(k',x) P_L \Delta^{<,>}(k-k',x) \,. 
\label{self1}
\eeq 
Explicit analytical forms for $\Sigma^{<,>}$ can be found in ref.~\cite{HKR2}. Numerical results for the total asymmetry and left chiral current density are shown in Fig.~\ref{fig:fermions}. For various parameters we used (in units of $|{\bf k}|$) $\tau_w=2$, $\Delta\theta=\frac12$, $T=10$, $m_q=10$, $m_{\phi}=5$ and $y=0.3$ (0 in the free case) and the initial conditions at $t \to -\infty$ were set to to thermal equilibrium: $f^h_\pm \to 1/(e^{\pm \beta \omega_k}+1)$ and $f^h_{1,2} \to 0$. In the collisionless case no total asymmetry is created and the L-chiral asymmetry oscillates with a constant amplitude after the transition. In the interacting case a temporary nonzero total asymmetry is created and both the total asymmetry and the amplitude of the oscillating L-chiral asymmetry decay as a function of time. Such a simple picture might be used to model the baryon asymmetry generation during a second order (or cross-over) electroweak phase transition, for example in models where the expansion rate of the universe is highly accelerated during the EWPT.

\section{Scalar fields}
\label{sec:scalar}

Our method can also be applied to the bosonic fields with similar results. In the case of the scalar fields we wish to study the correlator:
$i\prop^<(u,v) \equiv \langle  \phi(v)\phi(u)\rangle \equiv 
{\rm Tr}\left\{\hat \rho \ \phi(v)\phi(u)\right\}$.
The derivation of the KB-equations for $\prop^<$ proceeds analogously to the fermionic case~\cite{HKR3}; the final result in the mixed representation is
\begin{equation}
 \prop_0^{-1}\prop^<
  -  e^{-i\Diamond}\{ \Pi_H \}\{ \prop^< \}
  -  e^{-i\Diamond}\{ \Pi^< \}\{ \prop_H \}
  = {\cal C}_{\rm coll} \,,
\label{ScalarDynEqMix}
\end{equation}
with
\begin{equation}
{\cal C}_{\rm coll} = -i e^{-i\Diamond}
                             \left( \{\Gamma\}\{\prop^<\} -
                                    \{\Pi^<\}\{{\cal A_\phi}\}\right)\,,
\label{Scalarcollintegral}
\end{equation}
where we already used the connection ${\cal A}_\phi = \sfrac{i}{2}(\Delta^> + \Delta^<)$ and defined $\Gamma \equiv \sfrac{i}{2}(\Pi^>+\Pi^<)$. The self energy functions are again computable from the 2PI-generating function: $\Pi^{<,>}(u,v) \equiv -i\delta \Gamma_2[\prop ]/\delta \prop^{>,<}(v,u)$. Of course the explicit forms of $\prop_0$ as well as the interactions depend on the model. Similarly to the fermionic case, we are interested in effects caused by a space-time varying mass function, so we take
\begin{equation}
{\cal L} = \sfrac{1}{2} (\partial_\mu \phi)^2 -\sfrac{1}{2} m(x)^2 \phi^2 + {\cal L}_{\rm int} \,,
\label{ScalarfreeLag1}
\end{equation}
where ${\cal L}_{\rm int}$ is some yet to be defined interaction term.  The inverse free propagator corresponding to Eq.~(\ref{ScalarfreeLag1}) in the mixed representation then is
\begin{equation}
\prop_0^{-1} \equiv 
      k^2 - \sfrac{1}{4} \partial^2 + ik\cdot\partial 
      - m^2 e^{-\frac{i}{2}\partial^m_x \cdot \partial^\prop_k} \,,
\label{freeprop}
\end{equation}
where the $\partial^m_x$-derivative always acts on the mass term and the $\partial^\prop_k$-derivative to $\prop^<$. We again consider the free theory first, and also expand the $\prop_0$ operator to the lowest nontrivial order in gradients.  Because the operator is complex, we find two independent equations:
\begin{eqnarray}
 \Big(k_0^2 -{\bf k}^2 - m^2(t) - \sfrac{1}{4} \partial_t^2 \Big)i\prop^<(k,t)
      &=& 0
\label{scalarKG_Eq_HOM2}\\
 k_0\partial_t i\prop^<(k,t) &=& 0\,.
\label{scalarKG_Eq_HOM3}
\end{eqnarray}
Unlike with fermions, the noninteracting scalar KB-equations can not be divided into algebraic constraints and dynamic equations because both equations (\ref{scalarKG_Eq_HOM2}-\ref{scalarKG_Eq_HOM3}) contain time derivatives. Yet a similar shell structure does emerge. Let us first assume that $k_0 \neq 0$. Then\footnote{Note that including collisions at this stage would destroy the singular phase space structure. Neglecting them here corresponds to the usual quasiparticle approximation.} Eq.~(\ref{scalarKG_Eq_HOM3}) requires that $\partial_t i\prop^< = 0$ at all times and so one must also have $\partial_t^2 i\prop^< = 0$. Substituting this back to Eq.~(\ref{scalarKG_Eq_HOM2}) gives
\begin{equation}
\left(k_0^2 -{\bf k}^2 -m^2(t)\right)i\prop^<_{\rm m-s} = 0\,.
\label{alkepraeq1}
\end{equation}
This equation has the spectral mass-shell solution parametrized by $t$:
\begin{equation}
i\prop^<_{\rm m-s}(k_0,|{\bf k}|,t) = 
    2\pi\,{\rm sgn}(k_0)f_{s_{k_0}}(|{\bf k}|,t)
    \delta\big(k_0^2 -{\bf k}^2 -m^2(t)\big)\,, 
\label{SpecSolHOM}
\end{equation}
where $s_{0}\equiv{\rm sgn}(k_0)$. However, the point $k_0=0$ is special: if we first set $k_0=0$, then equation (\ref{scalarKG_Eq_HOM3}) is identically satisfied, but Eq.~(\ref{scalarKG_Eq_HOM2}) becomes $\partial_t^2 \prop^< = - 4 \omega_{{\bf k}}^2(t) \prop^<$. For a generic time-varying mass this implies a corresponding spectral solution for $k_0=0$:
\begin{equation}
i\prop^<_{\rm  0-s}(k_0,|{\bf k}|,t) = 2\pi\,f_c(|{\bf k}|,t)\delta(k_0)\,,
\label{k0zerospec}
\end{equation}
where $f_c(|{\bf k}|,t)$ is some real-valued function. The most complete solution for a given momentum $|{\bf k}|$ is the combination of the solutions (\ref{SpecSolHOM}) and (\ref{k0zerospec}):  $\prop^< = \prop^<_{\rm m-s} + \prop^<_{\rm 0-s}$. The situation is now seen to be qualitatively equivalent to the case with fermions and we interpret analogously that the new $k_0=0$-solution (\ref{k0zerospec}) describes the quantum coherence between particles and antiparticles.

\subsection{Moment connection}

Let us again suppose that we do not have any prior restrictions on the momentum variable ${\bf k}$. It is then appropriate to use the $n$-th moments as the weighted density functions for the problem:
\begin{equation}
\rho_n(|{\bf k}|,t) = \int \frac{{\rm d} k_0}{2\pi}\;k_0^n\,
i\prop^<(k_0,|{\bf k}|,t)\,.
\label{n-moment}
\end{equation}
When one computes three lowest moments using the complete shell structure of $\prop^<$ one immediately finds:
\begin{eqnarray}
\rho_0 &=& \sfrac{1}{{2\omega_{{\bf k}}}}(f_+ - f_-) + f_c 
\nonumber\\
\rho_1 &=& \sfrac{1}{2}(f_+ + f_-)
\nonumber\\[1mm]
\rho_2 &=& \sfrac{{\omega_{{\bf k}}}}{2}(f_+ - f_-)\,.
\label{rho-f_HOM}
\end{eqnarray}
All higher moments can be trivially related to these three lowest ones.
As was the case with fermions, these relations are only valid to the lowest order in mass gradients. However, when used as an {\em ansatz} in the complete set of equations (up to all order in gradients in the flow term), they lead to a closure and a solvable set of equations for the moment functions $\rho_n$.

\begin{figure}
\centering
\includegraphics[width=10cm]{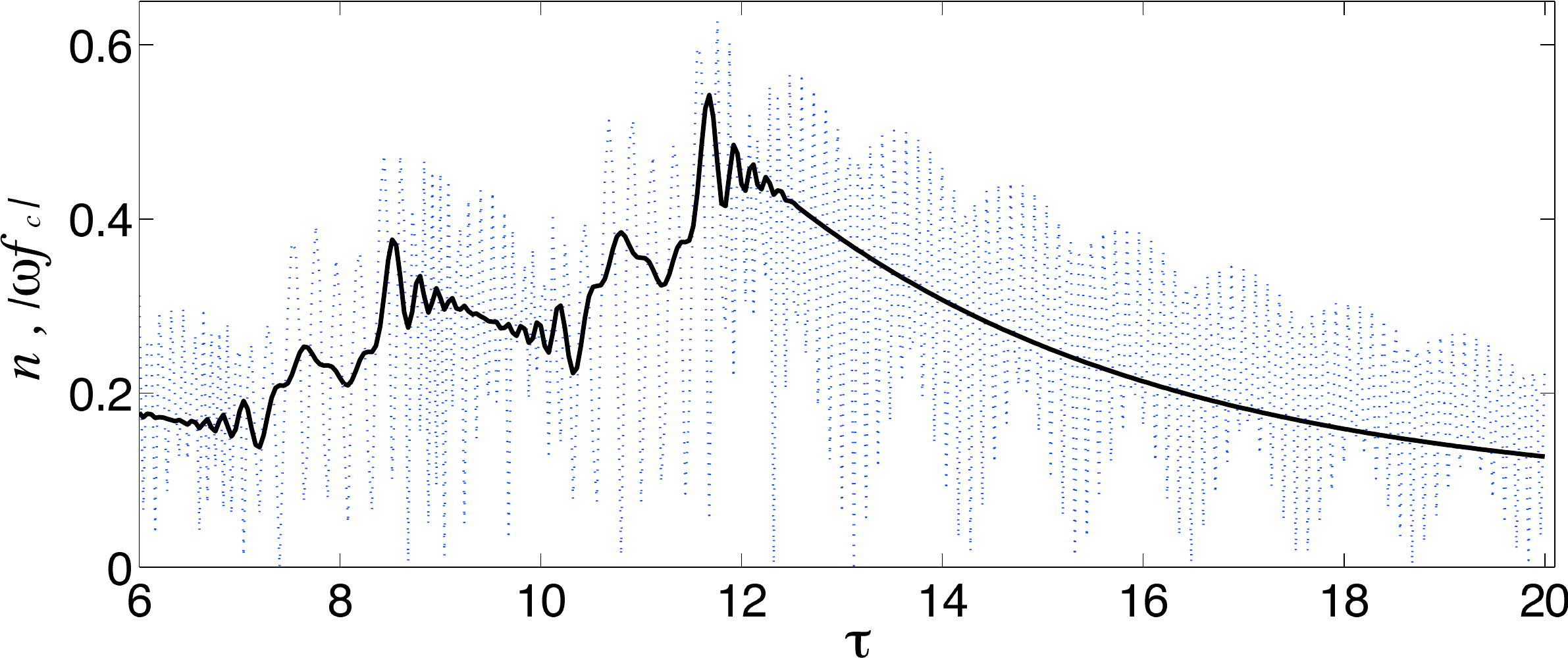}
    \caption{\label{fig:numberdensity2} Shown is the number density $n_{{\bf k}}$ (thick solid line) and the coherence function $f_c(|\bf k|)$ (dotted(blue) line). For the driving mass function Eq.~(\ref{massfunction}) we used the parameters $m_0 = T$,  $A = 1.5T$ and $B=0.1T$. For other parameters we used $|{\bf k}| = 0.6 T$, $y = 1$, $m_\psi=0.1 T$ and $\omega_\varphi = 0.1 T$, where temperature $T$ sets the scale. At $\tau \equiv \omega_\varphi t = 0$ the system was in the adiabatic vacuum, and the driving mass term smoothly set to constant $m = 2.5 \,T $ after $\tau > 4 \pi$.}    
\end{figure}

\subsection{Evolution equations with collisions}

When the cQPA ansatz is inserted to the full KB-equation (\ref{ScalarDynEqMix}) (again dropping terms with $\prop_H$ and $\Pi_H$), and the resulting equations are integrated over the energy, and the projections of the collision terms onto $k_0=0$ shell are resummed, one finds the following closed set of equations~\cite{Matti}:
\begin{eqnarray}
\sfrac{1}{4} \partial_t^2 \rho_0 + \omega_{{\bf k}}^2 \rho_0 - \rho_2 &=& -\sfrac{1}{{2\omega_{\bf k}}}\Gamma_m\,\partial_t\rho_0
\nonumber\\
\partial_t\rho_1 &=& -\sfrac{1}{{\omega_{{\bf k}}}}\Gamma_m \left(\rho_1 - \rho_{1,{\rm
      eq}}\right)
\nonumber\\
\partial_t\rho_2 - \sfrac{1}{2} \partial_t(m^2) \rho_0 &=& -\sfrac{1}{{\omega_{{\bf k}}}}\Gamma_m \left(\rho_2 - \rho_{2,{\rm
      eq}}\right)\,.
\label{rho_Eq_Coll2}
\end{eqnarray}
We illustrate the solution of Eqs.~(\ref{rho_Eq_Coll2}) in the case where the scalar mass term is an oscillatory function coming from a coupling of $\phi$ to an oscillating background field $\varphi$ (the inflaton):
\beq
m^2(t) = |m_0 + A \cos(2\omega_\varphi t) + i B\sin(2 \omega_\varphi t)|^2 \,,
\label{massfunction}
\eeq
Physically this situation could model the particle production after inflation  (preheating). Interactions were modelled by decays and inverse decays caused by the interaction with a thermalized fermion background: ${\cal L}_{\rm int} = g \phi \bar\psi \psi$, where $\psi$ is some generic thermalized background fermion field.  As the inflaton oscillates around its minimum, $\phi$-fields are excited out from the temporally varying classical background and the number density increases in each oscillation cycle.  Also the magnitude of the coherence function increases at each oscillation, although the decohering effect of collisions is clearly visible between the oscillation periods.  After the driving mass terms is set to constant, a phase of smooth decoherence and thermalization is observed.

\section{Conclusions}
\label{sec:conclu}

We have introduced a novel approximation for the Kadanoff-Baym equations for the fermionic and bosonic 2-point functions. Our approach makes use of newly discovered extended singular phase space structure for these functions in the free theory limit in two simple geometries: the spatially homogeneous and stationary planar symmetric cases. When applied to the full KB-equations, this {\em coherent quasiparticle approximation} (cQPA) leads to equations of motion for a density matrix (or a set of moment functions for scalar fields), that can describe nonlocal coherent evolution of the state in the presence of hard collisions. In the homogeneous case the new shell solutions characterize the particle-antiparticle correlations  and in the planar symmetric case correlations between reflecting states moving on opposite directions. Our original, and still principal motivation for developing this formalism was the desire to find a quantitative description for the electroweak baryogenesis proceeding through the CP-violating reflection mechanism. While we yet have to complete this program, we have already demonstrated the use of our formalism in a number of other interesting problems; for example the homogeneous time dependent problem of particle production in the early universe at the end of inflation or some phase transition. This formalism can be used to model homogeneous baryogenesis as well. Yet another interesting research direction is to use the cQPA to better understand the nature of the flavour coherence in the case of neutrino mixing and in particular in connection with the supernova neutrinos.


\section*{References}
\begin{thebibliography}{9}

\bibitem{BG} 
  A.~Cohen, D.~Kaplan, and A.~Nelson, 
    \emph{Progress in electroweak baryogenesis}, in 
    \emph{43}{1994}{27} [\hepph{9302210}];\\
  V.A.~Rubakov and M.E.~Shaposhnikov, 
  Usp.~Fiz.~Nauk {\bf 166} (1996) 493,
  Phys.~Usp.~{\bf 39} (1996) 461.

\bibitem{ClassForce}
  M.~Joyce, T.~Prokopec and N.~Turok,
  Phys.\ Rev.\ D {\bf 53}, 2958 (1996); 
  Phys.\ Rev.\ Lett.\  {\bf 75}, 1695 (1995);
  [Erratum-ibid.\  {\bf 75}, 3375 (1995)];
  Phys.\ Rev.\ D {\bf 53}, 2930 (1996).

\bibitem{ClassForceomat}
  J.~M.~Cline, M.~Joyce and K.~Kainulainen,
  JHEP {\bf 0007} (2000) 018; Phys.\ Lett.\ B {\bf 417} (1998) 79,
  [Err.-ibid.\ B {\bf 448} (1999) 321];
  J.~M.~Cline and K.~Kainulainen,
  {Phys.\ Rev.\ Lett.\ }  {\bf 85} (2000) 5519.

\bibitem{SemiClassSK}
  K.~Kainulainen, T.~Prokopec, M.~G.~Schmidt and S.~Weinstock,
  JHEP {\bf 0106}, 031 (2001);
  Phys.~Rev.~D{\bf 66} (2002) 043502;
  T.~Prokopec, M.~G.~Schmidt and S.~Weinstock,
  Annals Phys.~{\bf 314}, 208 (2004);
  Annals Phys.~{\bf 314}, 267 (2004).

\bibitem{PSW}
  T.~Prokopec, M.G.~Schmidt and S.~Weinstock,
  Annals Phys.~{\bf 314}, 208 (2004);
  Annals Phys.~{\bf 314},267 (2004).

\bibitem{baym} See G.~Baym, these proceedings. 

\bibitem{brandenberger}
  J.~H.~Traschen and R.~H.~Brandenberger,
  Phys.~Rev.~D {\bf 42} 2491 (1990).

\bibitem{generic_therm} J.~Berges, S.~Borsanyi, J.~Serreau, 
  Nucl.~Phys.~B {\bf 660}, 51 (2003). 
  
\bibitem{HKR1}
  M.~Herranen, K.~Kainulainen and P.~M.~Rahkila,
  Nucl.\ Phys.\  B {\bf 810} (2009) 389.

\bibitem{HKR2}
  M.~Herranen, K.~Kainulainen and P.~M.~Rahkila,
  JHEP {\bf 0809} (2008) 032.

\bibitem{HKR3}
  M.~Herranen, K.~Kainulainen and P.~M.~Rahkila,
  JHEP {\bf 0905} (2009) 119.
   
\bibitem{HKR4}
  M.~Herranen, K.~Kainulainen and P.~M.~Rahkila,
  Nucl.\ Phys.\  A {\bf 820} (2009) 203C.
   
\bibitem{HKR5} M.~Herranen, K.~Kainulainen and P.M.~Rahkila, in progress. 

\bibitem{thooft}
  G.~'t Hooft, Phys.\ Rev.\ Lett.\  {\bf 37} (1976) 8.

\bibitem{RummuMoore} G.D.~Moore and K.~Rummukainen, Phys.~Rev.~D61 (2000) 105008; D.~B\"odeker, G.D.~Moore and K.~Rummukainen, Nucl.~Phys.~Proc.~Suppl.~83:583-585,2000. 

\bibitem{Schwinger-Keldysh}
  J.~S.~Schwinger, 
  J.~Math.~Phys.~{\bf 2} 407 (1961); \\ 
  L.~V.~Keldysh, 
  Zh.~Eksp.~Teor.~Fiz.~{\bf 47} 1515 (1964) [Sov.\
  Phys.~JETP {\bf 20} 1018 (1965)]. 

\bibitem{CTP}
  K.~c.~Chou, Z.~b.~Su, B.~l.~Hao and L.~Yu, 
  Phys.~Rept.~{\bf 118}, 1 (1985); 
  E.~Calzetta and B.~L.~Hu, Phys.\ Rev.\  D{\bf 37} (1988) 2878.

\bibitem{kaksPI}
  J.~M.~Cornwall, R.~Jackiw and E.~Tomboulis, 
  Phys.\ Rev.\  D {\bf 10} (1974) 2428.

\bibitem{KB}
  L.~Kadanoff and G.~Baym, ``Quantum Statistical Mechanics''
  {\it Benjamin, New York (1962)}.

\bibitem{Prokopec_partnumber} B.~Garbrecht, T.~Prokopec and M.G.~Schmidt
Eur.~Phys.~J.~C38, 135 (2004).

\bibitem{CKV}
  J.~M.~Cline, K.~Kainulainen and A.~P.~Vischer,
  Phys.\ Rev.\  D {\bf 54} (1996) 2451.

\bibitem{Henning} P.~Henning, Phys.~Rept.~{\bf 253} (1995); G. Mahan, Phys.~Reps.~145 (1987) 251.

\bibitem{Matti} M.~Herranen, PhD Dissertation, University of Jyv\"askyl\"a,   [arXiv:0906.3136 [hep-ph]].


\end{thebibliography}
\end{document}